\documentclass[
twocolumn,secnumarabic
,amssymb, amsmath,nobibnotes, aps, prl]{revtex4}
\usepackage{graphicx}
\usepackage{tabularx}
\usepackage{bm}%
\expandafter\ifx\csname package@font\endcsname\relax\else
 \expandafter\expandafter
 \expandafter\usepackage
 \expandafter\expandafter
 \expandafter{\csname package@font\endcsname}%
\fi

\begin{document}

\title{Quantized Noncommutative Geometry from Multitrace Matrix Models}

\author{Badis Ydri$^{a}$, Ramda Khaled$^{b}$, Cherine Soudani$^{c}$}
\affiliation{$^{a}$Department of Physics, Badji-Mokhtar Annaba University,\\
 Annaba, Algeria\\
  $^{b}$Ecole Normale Sup\'erieure Messaoud-Zeghar,\\
  Setif, Algeria\\
$^{c}$Physics Department, Hamma-Lakhdar El Oued University,\\
El Oued, Algeria.
}

\begin{abstract}
In this article the geometry of quantum gravity is quantized in the sense of being noncommutative (first quantization) but it is also quantized in the sense of being emergent (second quantization). A new mechanism for quantum geometry is proposed in which noncommutative geometry can emerge from "one-matrix multitrace scalar matrix models"  by probing the statistical physics of commutative phases of matter. This is  in contrast to the usual mechanism in which noncommutative geometry emerges from "many-matrix singletrace Yang-Mills matrix models" by probing the statistical physics of noncommutative phases of gauge theory.  In this novel scenario quantized geometry emerges in the form of a transition between the two phase diagrams of the real quartic matrix model and the noncommutative scalar phi-four field theory. More precisely, emergence of the geometry is identified here with the emergence of the uniform-ordered phase and the corresponding commutative (Ising) and noncommutative (stripe) coexistence lines. The critical exponents and the Wigner's semicircle law are used to determine the dimension and the metric respectively. Arguments from the saddle point equation, from Monte Carlo simulation and from the matrix renormalization group equation are provided in support of this scenario.

\end{abstract}

\maketitle
\tableofcontents
\section{Introduction: Noncommutative geometry and matrix models}
Noncommutative field theory is quantum field theory defined on a noncommutative spacetime which is primarily characterized by a minimum length scale believed to be of the order of the Planck scale. This minimum length scale appears in a very natural way in almost all theories of quantum gravity such as string theory and loop quantum gravity. Noncommutative spacetime is also conjectured to be characterized by a discrete structure which is naturally captured by operator algebras and matrix models. Hence, a noncommutative field theory and its underlying noncommutative spacetime, similarly to the theory of quantum mechanics, are characterized by the two fundamental features:
\begin{enumerate}
\item A Planck-like length scale (noncommutativity parameter). A noncommutative structure should then be understood as "first quantization" of the commutative geometry.
\item A discrete geometry which becomes continuous at large distances (commutative limit). The operator algebras and in particular the underlying matrix models (together with their Feynman path integrals) should then be understood as providing "second quantization" of the geometry.
\end{enumerate}
These matrix models capture then quantum gravitational fluctuations around the "classical" noncommutative structures and the commutative limit is akin to the semi-classical limit. This means in particular that noncommutative spacetimes, similarly to phase spaces of classical mechanics, and the corresponding classical gravity are constrained to be defined on Poisson/symplectic manifolds. 

Noncommutative field theory and the underlying noncommutative spacetime require for their description the language of noncommutative geometry \cite{connes,Connes:1996gi}.  In this context the geometry is in a precise sense emergent (here from operator algebras) given in terms of a spectral triple $({\cal A}, \Delta, {\cal H})$ rather than in terms of a set of points. The algebra ${\cal A}$ is typically an operator operator represented on the Hilbert space ${\cal H}$  whereas $\Delta$ is the Laplace operator which defines the metric aspects of the geometry. We really should think of the algebra ${\cal A}$ as defining the topology whereas the Laplacian $\Delta$ defines the metric. And in the non-perturbative formulation of the spectral triple  $({\cal A}, \Delta, {\cal H})$ as a path or functional integral we require the language of matrix models.

The most compelling physical framework in which noncommutative field theory, noncommutative geometry and matrix models appear very naturally can be found in string theory \cite{Seiberg:1999vs,Connes:1997cr}.

Among the first studied examples of a noncommutative field theory is a scalar field theory on a Groenewold-Moyal-Weyl spacetime \cite{weyl,Moyal:1949skv2,Groenewold:1946kpv2} which was considered in \cite{Filk:1996dm}. This theory remains one of the most important examples in noncommutative geometry, noncommutative field theory and matrix models for two essential reasons. First, because of the rich phase structure characterizing a noncommutative scalar field theory which is of great implications for both particle physics, e.g. spontaneous symmetry breaking and condensed matter systems, e.g. exotic phases of matter and quantum phase transitions \cite{Minwalla:1999px}. Second, also because the underlying Groenewold-Moyal-Weyl spacetime is the most important noncommutative spacetime as it plays for Poisson manifolds (via Darboux theorem) the same role played by flat Minkowski spacetime for curved manifolds, i.e. Darboux theorem is effectively the equivalence principle in this case as noted for example in \cite{Lee:2010zf1,Blaschke:2010ye}.

A noncommutative scalar field theory is given by an action functional which can be split in the usual fashion into a kinetic part and a potential term. This action functional is defined non-perturbatively by means of a matrix/operator model (after proper regularization and Euclidean rotation). The potential term represents the algebraic structure, i.e. the algebra, the inner product, the Hilbert space, etc and as such it defines in some sense the topological aspects of the noncommutative space.  The kinetic term on the other hand, as it involves a Laplacian operator $\Delta$, plays a pivotal role in defining the metric aspects of the geometry following Fr\"{o}hlich and Gaw\c{e}dzki \cite{FroehlichGawedzki} (or Connes \cite{connes} for spin geometry).

Thus, a noncommutative scalar field theory on some underlying noncommutative spacetime (assumed to be a Moyal-Weyl spacetime for concreteness) is given generically by a matrix model of the form
\begin{eqnarray}
S=a {\rm Tr}\Phi\Delta\Phi+{\rm Tr}V(\Phi)~,~\Delta(..)= \alpha [X_a,[X_a,..]].\label{matrix}
\end{eqnarray}
For simplicity, we will only consider here a one-matrix real model, i.e. we have a single Hermitian scalar field $\Phi=\Phi^{\dagger}$. In generic cases  the trace ${\rm Tr}$ is over an infinite dimensional Hilbert space ${\cal H}$ and the scalar field $\Phi$ is an operator, i.e. an element of the algebra ${\cal A}$. If we further assume that an appropriate regularization is employed then the trace ${\rm Tr}$ becomes cutoff and the operator $\Phi$ becomes an $N\times N$ matrix. For example, we can generally regularize the Moyal-Weyl spacetime using noncommutative tori  \cite{rieffel,Ambjorn:2000cs} or fuzzy spheres \cite{Hoppe,Madore:1991bw}.

The operators $X_a$ entering the definition of the Laplacian operator $\Delta$ are, in most important cases, the coordinate operators which define operationally the noncommutative spacetime. For example, the Moyal-Weyl spacetime is precisely defined in terms of the operators $X_a$ by the Heisenberg relations 
\begin{eqnarray}
[X_a,X_b]=i\Theta_{ab}~,~\Theta_{ab}={\rm constant}.
\end{eqnarray}
The noncommutativity parameter $\Theta_{ab}$ for general noncommutative spaces is a function of the operators $X_a$, i.e. $\Theta_{ab}\equiv \Theta_{ab}(X)$.

We will assume now that an Euclidean rotation and a regularization are employed and as such we are indeed dealing with a one-matrix real matrix model.  The action functional is still given by equation (\ref{matrix}) but $\Phi$ is now a Hermitian $N\times N$ matrix. We will further assume that the interaction term is given by a quartic phi-four potential, viz
\begin{eqnarray}
V(\phi)=b \Phi^2+c \Phi^4.\label{phi-four}
\end{eqnarray}
The phase structure of this matrix model was calculated non-perturbatively mostly in two dimensions on the fuzzy sphere (but also on the noncommutative torus) using mostly the Monte Carlo method \cite{GarciaFlores:2009hf,Martin:2004un,Panero:2006bx,Das:2007gm,Ydri:2014rea}. In this case the Laplacian is given by $\Delta(..)=  [L_a,[L_a,..]]$ where $L_a$ are the generators of $SU(2)$ in the spin irreducible representation $j=(N-1)/2$.  It is believed that the phase diagram of noncommutative phi-four theory in any dimension and on any noncommutative background is identical in features to the phase diagram of the potential (\ref{phi-four}) on the fuzzy sphere.   

In the remainder of this article we will present the multitrace matrix model and the saddle point equation (section $2$), the phase diagram and the Monte Carlo algorithm (section $3$), the Wilsonian matrix renormalization group equation (section $4$), the proposal of "emergent geometry from random multitrace matrix model" (section $5$) and then we conclude with a summary and a brief outlook (section $6$).
\section{The multitrace matrix model and the saddle point equation}
The multitrace approach to noncommutative scalar field theories was initiated in \cite{O'Connor:2007ea,Saemann:2010bw} on the fuzzy sphere.  For an earlier approach see \cite{Steinacker:2005wj} and for a similar non-perturbative approach see \cite{Polychronakos:2013nca,Tekel:2014bta,Nair:2011ux,Tekel:2013vz} and \cite{Tekel:2015uza,Tekel:2015zga,Subjakova:2020haa}.

In this approach we diagonalize the Hermitian scalar matrix $\Phi$ as $\Phi=U M U^{\dagger}$ and then integrate over the unitary matrix $U$ using the group theoretic structure and properties of $SU(2)$ and $SU(N)$ extensively. In effect,  in this approach the kinetic term is expanded while the potential term is treated exactly. This is in fact a hopping-parameter-like expansion. The end result is to convert the kinetic term into a multitrace matrix model, which to the lowest non-trivial order, is of the form
\begin{eqnarray}
&&\int dU \exp\bigg(a{\rm Tr}[L_a,UMU^{\dagger}]^2-{\rm Tr}V(M)\bigg)\nonumber\\
&=&\exp\bigg(B{\rm Tr}M^2+C{\rm Tr}M^4+D({\rm Tr}M^2)^2+B^{\prime}({\rm Tr}M)^2\nonumber\\
&+&C^{\prime}{\rm Tr}M{\rm Tr}M^3+D^{\prime}({\rm Tr}M)^4+A^{\prime}{\rm Tr}M^2({\rm Tr}M))^2+...\bigg).\nonumber\\
\end{eqnarray}
The parameters $B$ and $C$ have shifted values with respect to the original parameters of the potential $b$ and $c$ respectively while the primed parameters and $D$ are have purely quantum values coming from the hopping-parameter-like expansion of the kinetic term.

The basic statement is that the Laplacian operator on the fuzzy sphere (which is the example considered here) is exactly equivalent to this multitrace matrix model with these particular value of the coefficients. And each Laplacian operator, i.e. any other noncommutative space comes with its own set of multitrace coefficients. 

\begin{figure}[htbp]
\begin{center}
\includegraphics[width=9.0cm,angle=-0]{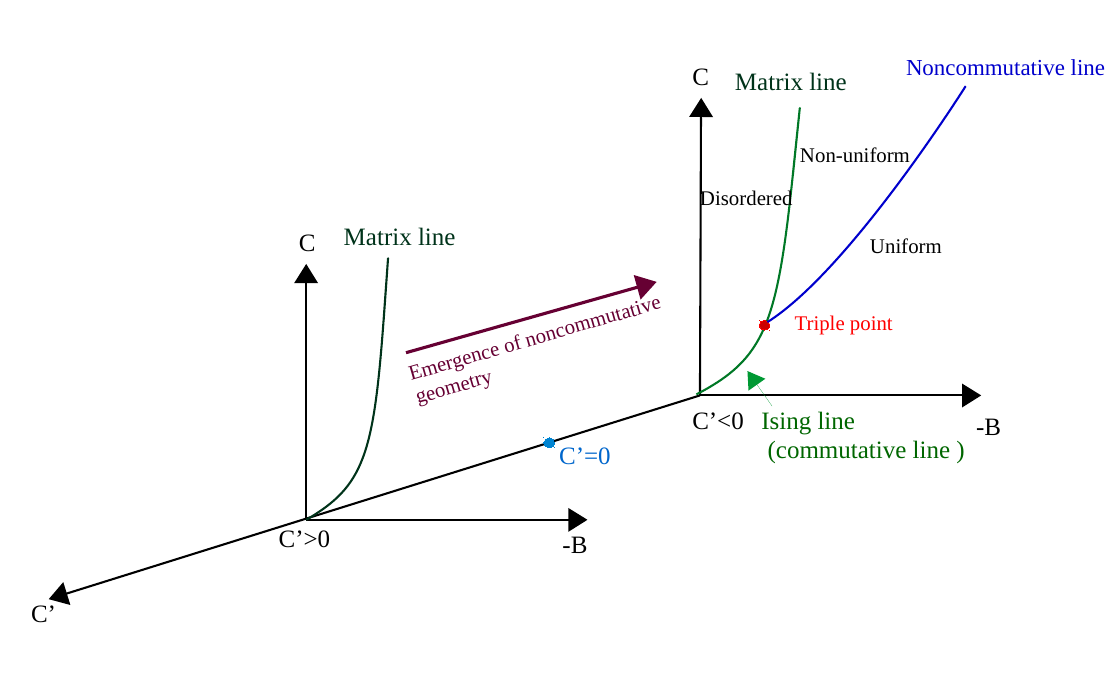}
\caption{Emergent geometry  as a transition between the two phase diagrams of the real quartic matrix model and the noncommutative scalar phi-four field theory. Emergence of the geometry here is identified with the emergence of the uniform-ordered phase.}\label{emergent_geometry}
\end{center}
\end{figure}

It has been shown in   \cite{Ydri:2015zsa,Ydri:2017riq} that the essential features of the phase diagram of noncommutative phi-four theory in two dimensions can be captured by a truncated multitrace matrix model depending on the cubic moment ${\rm Tr}M^3$, i.e. a multitrace matrix model given simply by the potential

\begin{eqnarray}
V_{\rm trunc}=
B{\rm Tr}M^2+C{\rm Tr}M^4+C^{\prime}{\rm Tr}M{\rm Tr}M^3.\label{multitrace}
\end{eqnarray}
Naturally, stability requirement of this multitrace matrix model constrains the range of the quartic coupling $C$ in a particular way. The statistical physics of this multitrace matrix model interpolates between the statistical physics of the noncommutative field theory (\ref{matrix})+(\ref{phi-four}) and the statistical physics of the real quartic matrix model given by 
\begin{eqnarray}
V_{\rm pure}=
B{\rm Tr}M^2+C{\rm Tr}M^4.\label{purematrix}
\end{eqnarray}
Indeed, the random multitrace matrix model (\ref{multitrace}) lies in the universality class of noncommutative phi-four theory for $C^{\prime}$ negative whereas it lies in the universality class of the real quartic matrix model for $C^{\prime}$ positive. See figure (\ref{emergent_geometry}).

This idea and its generalization to more general multitrace matrix models and other noncommutative spaces is the underpinning of the proposal of emergent geometry from random multitrace matrix models which was originally laid down in \cite{Ydri:2020efr,Ydri:2017riq,Ydri:2016daf,Ydri:2015zsa} using different methods (renormalization group equation, Monte Carlo algorithm and large $N$ saddle point method).

The above random multitrace matrix model  should be thought of as a generalization of the discretization of random Riemannian surfaces with regular polygons such as dynamical triangulation \cite{DiFrancesco:1993cyw,Zarembo:1998uk}. Indeed, and as we will discuss shortly, random multitrace matrix models can sustain emergent geometry as well as growing dimensions and topology change. For example, the multitrace matrix model (\ref{multitrace}) works in two dimensions but also it works away from two dimensions where it can generate a large class of spaces, such as fuzzy projective space $\mathbb{C}{\mathbb P}_N^n$ \cite{Balachandran:2001dd}, which admit finite spectral triples that can be captured by the multitrace term ${\rm Tr}M {\rm Tr}M^3$.

In terms of the collapsed or scaled parameters  $\tilde{B}=B/N^{3/2}$, $\tilde{C}=C/N^2$ and $\tilde{C}^{\prime}=C^{\prime}/N$ the multitrace matrix model (\ref{multitrace}) is rewritten (with the scaled field $\tilde{M}=N^{1/4}M$) as
\begin{eqnarray}
V_{\rm trunc}=
N\tilde{B}{\rm Tr}{M}^2+N\tilde{C}{\rm Tr}{M}^4+\tilde{C}^{\prime}{\rm Tr}{M}{\rm Tr}{M}^3.\label{multitrace1}
\end{eqnarray}
In the large $N$ limit the saddle point equation controls the statistical quantum physics of the theory. The saddle point equation and the corresponding  free energy read explicitly
\begin{eqnarray}
&&2\tilde{B}\lambda+4\tilde{C}\lambda^3+\tilde{C}^{\prime}{m}_3+3\tilde{C}^{\prime}{m}_1\lambda^2\nonumber\\
&-&2\int d\lambda^{\prime}\rho(\lambda^{'})\frac{1}{\lambda-\lambda^{\prime}}=0.\label{spe}
\end{eqnarray}
\begin{eqnarray}
\frac{E}{N^2}&=&\int d\lambda \rho(\lambda)(\tilde{B}\lambda^2+\tilde{C}\lambda^4)+\tilde{C}^{\prime}\int d\lambda\rho(\lambda)\int d\lambda^{\prime}\rho(\lambda^{\prime})\lambda^{\prime 3}\nonumber\\
&-&\frac{1}{2}\int d\lambda\rho(\lambda)\int d\lambda^{\prime}\rho(\lambda^{\prime})\log (\lambda-\lambda^{\prime})^2.
\end{eqnarray}
The last term in both equations comes from the Vandermonde determinant which causes the eigenvalues to repel each other and become spread evenly around zero \cite{Brezin:1977sv}. In the saddle point equation (\ref{spe}) there appears also the moments $m_q=\int d\lambda\rho(\lambda)\lambda^q$ which modify the behavior of the real quartic matrix model through the multitrace coupling $C^{\prime}$.

The phase diagrams of the full matrix model (\ref{matrix})+(\ref{phi-four}) shares  with the phase diagram of the multitrace matrix model (\ref{multitrace}) the same essential features up to and including the triple point \cite{Ydri:2016daf}. 

The matrix model (\ref{matrix})+(\ref{phi-four}) without kinetic term ($a=0$) can be solved exactly to obtain a phase structure consisting of A) a disordered (symmmetric, one-cut, disk) phase and B) a non-uniform ordered (stripe, two-cut, anuulus) phase separated by a third order transition line \cite{Brezin:1977sv,Shimamune:1981qf}. This is the so-called matrix line and it corresponds to the matrix model fixed point of noncommutative field theory at infinite noncommutativity $\theta=\infty$ \cite{Bietenholz:2004as,Becchi:2003dg,Grosse:2003nw,Ydri:2013zya}.

The matrix model (\ref{matrix})+(\ref{phi-four}) with a kinetic term ($a\neq 0$) is much harder to solve but much more interesting. The phase diagram, in addition to the above two phases A) and B), involves a C) uniform (asymmetric, one-cut, Ising) ordered phase. The three phases meet at a triple point where the three co-existence curves intersect. This phase diagram is also observed in the truncated multitrace matrix model (\ref{multitrace}).

\begin{figure}[htbp]
\begin{center}
\includegraphics[width=9.0cm,angle=-0]{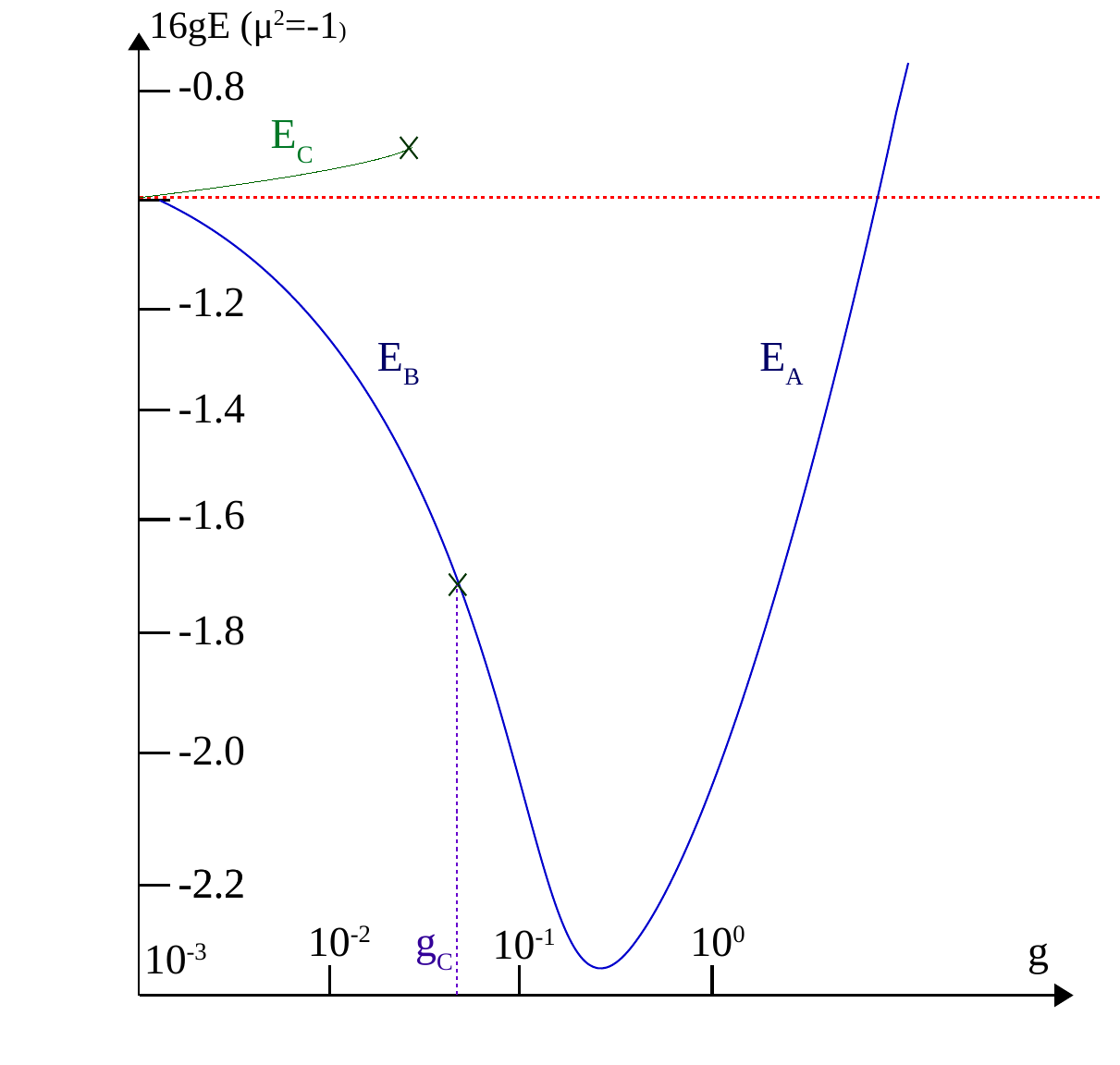}
\caption{The free energy of the real quartic matrix model (\ref{purematrix}) in the disordered, uniform-ordered non-uniform-ordered phases for $\mu^2=-1$. The coupling constants are identified as $g\equiv NC$ and $\mu^2\equiv 2B$. The uniform-ordered phase is metastable in this model \cite{Shimamune:1981qf}. }\label{free_energy}
\end{center}
\end{figure}

The main difference between the real quartic matrix model (\ref{purematrix}) from the one hand, and the noncommutative phi-four theory (\ref{matrix})+(\ref{phi-four}) on the other hand, lies in the fact that the uniform-ordered phase is stable in the latter. The three phases A), B) and C) in the real quartic matrix model are given explicitly by the following density eigenvalues  \cite{Shimamune:1981qf}
\begin{eqnarray}
A)~:~\rho(z)&=&\frac{1}{N\pi}(2Cz^2+B+C\delta^2)\sqrt{\delta^2-z^2},\nonumber\\
&&-B\leq -B_c=2\sqrt{NC}.
\end{eqnarray}
\begin{eqnarray}
B)~:~\rho(z)&=&\frac{2C|z|}{N\pi}\sqrt{(z^2-\delta_1^2)(\delta_2^2-z^2)},\nonumber\\
&&-B\geq -B_c=2\sqrt{NC}.
\end{eqnarray}
\begin{eqnarray}
C)~:~\rho(z)&=&\frac{1}{\pi N}(2C z^2 +2\sigma C z +B +2C\sigma^2+C\tau^2)\nonumber\\
&\times &\sqrt{((\sigma+\tau)-z)(z-(\sigma-\tau))},\nonumber\\
&&-B\geq -B_c=\sqrt{15}\sqrt{NC}.
\end{eqnarray}
The expressions of the various cuts $\delta$, $\delta_1$, $\delta_2$, $\sigma$ and $\tau$ in terms of the parameters of the model can be found in \cite{Shimamune:1981qf}. Following Shimamune it is then not very difficult to show that the free energy $E_C$ in the uniform-ordered phase is always higher than the free energies $E_A$ and $E_B$ in the disordered and non-uniform-ordered phases and hence the uniform-ordered phase is metastable in this model. In particular, the energies $E_B$ and $E_C$ are given explicitly by \cite{Shimamune:1981qf}
\begin{eqnarray}
E_B&=&-\frac{B^2}{4NC}+\frac{1}{4}\ln \frac{NC}{4B^2}-\frac{3}{8}\nonumber\\
E_C&=&-\frac{B^2}{4NC}+O(C^0)~,~C\longrightarrow 0.
\end{eqnarray}
In the limit $C\longrightarrow 0$ we have $E_B<E_C$ because of the logarithmic contribution. See figure (\ref{free_energy}).

Similarly, the main difference between the real quartic matrix model (\ref{purematrix}) from the one hand, and the multitrace matrix model (\ref{multitrace}) on the other hand, lies in the fact that the uniform-ordered phase is stable in the latter. This important fact can be seen in both the saddle point equation and the Monte Carlo algorithm.  For example, the saddle point equation (\ref{spe}) of the multitrace matrix model (\ref{multitrace}) can be solved exactly to determine the boundaries between the three phases A), B) and C) and the location of the triple point by following the remarkable work of \cite{Saemann:2010bw} who solved a much larger class of multitrace matrix models.

The important point to stress here is the fact that the uniform-ordered phase becomes stable within the multitrace matrix model (\ref{multitrace}), in contrast to the case of the pure matrix model (\ref{purematrix}), which can be shown using a mean-field-like analysis as follows.

First, by using the invariance of the partition function of the multitrace matrix model (\ref{multitrace}) under $M\longrightarrow (1+\epsilon)M$ we can derive the Schwinger-Dyson identity
\begin{eqnarray}
\frac{\langle V\rangle}{N^2}=\frac{1}{4}+\frac{\tilde{B}}{2}\langle \frac{1}{N}{\rm Tr}M^2\rangle.
\end{eqnarray}
Next we can use the classical configurations $M=0$,  $M=m_B\gamma$ and $M=m_C{\bf 1}$ deep inside the disordered, non-uniform-ordered and uniform-ordered phases to estimate the energies $E_A$, $E_B$ and $E_C$ respectively.

Thus, we calculate $m_C^2=-\tilde{B}/2(\tilde{C}+\tilde{C}^{\prime})$ which indicates that the model for $\tilde{B}<0$ is only stable if $\tilde{C}>-\tilde{C}^{\prime}$. The energy $E_C$ in the uniform configuration  is then estimated to be given by $E_C=(\tilde{C}+\tilde{C}^{\prime}-\tilde{B}^2)/4(\tilde{C}+\tilde{C}^{\prime})$. Similarly, we calculate $m_B^2=-\tilde{B}/2\tilde{C}$ using the fact that $\gamma^2={\bf 1}$.  The energy $E_B$ in the non-uniform configuration  is then estimated to be given by $E_B=(\tilde{C}-\tilde{B}^2)/4\tilde{C}$. It is not difficult to check that $E_B>E_C$ (and hence the uniform-ordered phase is much more stable than the non-uniform-ordered phase) if and only if $\tilde{C}^{\prime}<0$ and vice versa.

Remark that the effect of the multitrace term in the Ising phase is then only to shift the quartic coupling as $\tilde{C}\longrightarrow \tilde{C}+\tilde{C}^{\prime}$ whereas there is no effect in the stripe phase  from the multitrace term. Hence, the energies $E_B$ and $E_C$ in this mean-field-like approximation are estimated as follows 
\begin{eqnarray}
E_B&=&-\frac{B^2}{4NC}+\frac{1}{4}\ln \frac{N C}{4B^2}-\frac{3}{8}\nonumber\\
E_C&=&-\frac{B^2}{4N(C+C^{'})}+O((C+C^{'})^0)~,~C+C^{'}\longrightarrow 0.\nonumber\\
\end{eqnarray}
Since we can not reach the point $C=0$ (the model becomes unstable there) we can only take the limit  $C\longrightarrow -C^{'}$. In this case we have instead $E_C<E_B$.

\section{Monte Carlo simulation and the phase diagram}
As we have said, the noncommutative phi-four theory (\ref{matrix})+(\ref{phi-four}) and the random multitrace matrix model (\ref{multitrace}) share similar phase diagrams up to and including the triple point. Indeed, the multitrace matrix model (\ref{multitrace}) captures, in fact surprisingly very well, the  phase structure of noncommutative phi-four theory. This phase structure can be probed non-perturbatively using the Monte Carlo algorithm. 

A direct simulation of the noncommutative phi-four theory (\ref{matrix})+(\ref{phi-four}) is very involved for various physical and technical reasons  \cite{GarciaFlores:2009hf,Martin:2004un,Panero:2006bx,Das:2007gm,Ydri:2014rea}. Thus, the numerical results reported here are simply obtained by applying the Monte Carlo method or more precisely  the Metropolis algorithm to the multitrace matrix model (\ref{multitrace}) with the value $C^{'}=-N$ which is the value obtained for the multitrace interaction ${\rm Tr}M{\rm Tr}M^3$ in the multitrace expansion of the kinetic term in the noncommutative phi-four theory (\ref{matrix})+(\ref{phi-four}) on the fuzzy sphere \cite{O'Connor:2007ea,Saemann:2010bw}.

\begin{figure}[htbp]
\begin{center}
\includegraphics[width=10.0cm,angle=-0]{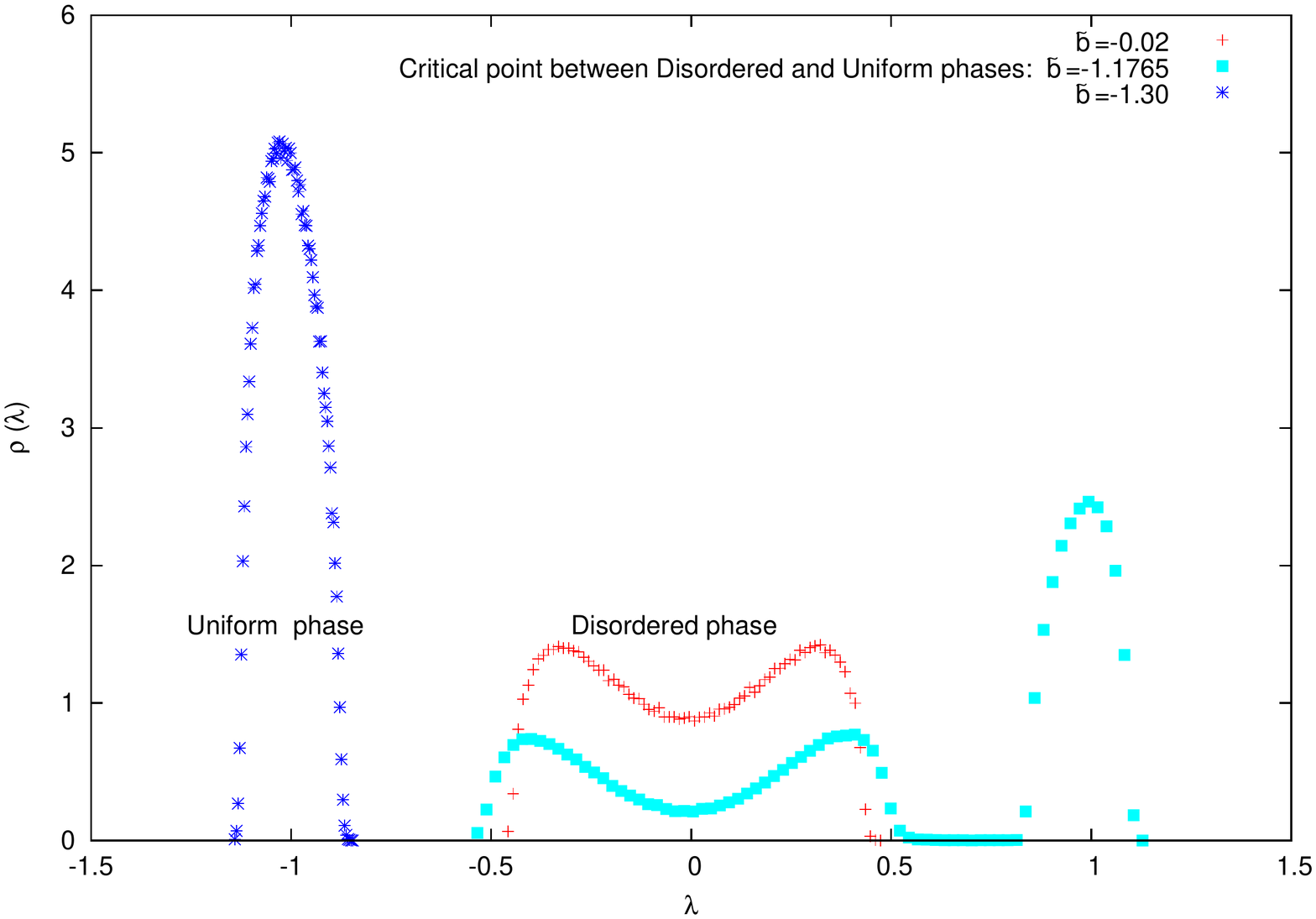}
\caption{The eigenvalue distribution as it transits between the disordered and uniform-ordered phases in the multitrace matrix model (\ref{multitrace}). }\label{commutative}
\end{center}
\end{figure}

\begin{figure}[htbp]
\begin{center}
\includegraphics[width=10.0cm,angle=-0]{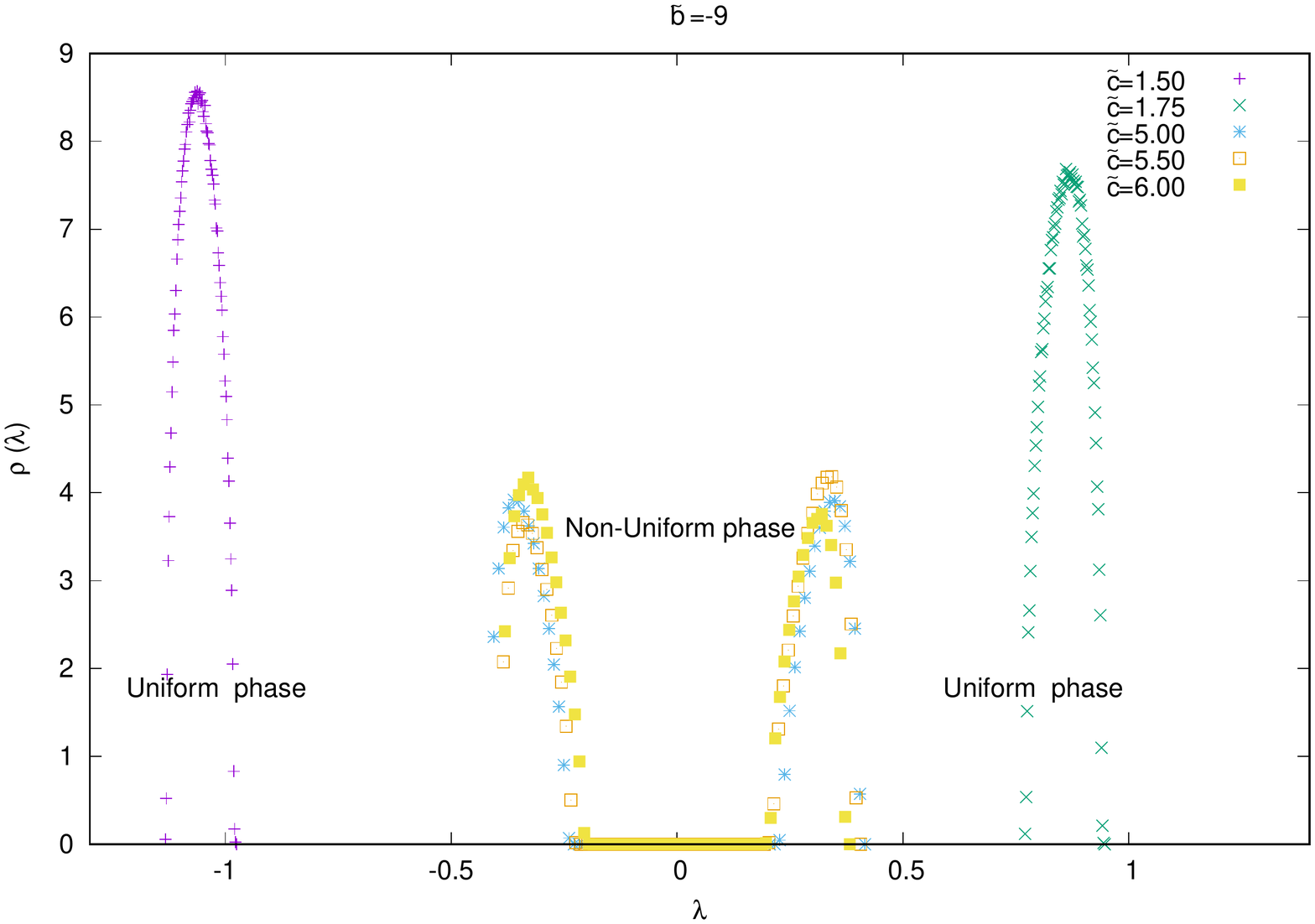}
\caption{The eigenvalue distribution as it transits between the uniform-ordered and non-uniform-ordered phases  in the multitrace matrix model (\ref{multitrace}). }\label{noncommutative}
\end{center}
\end{figure}

\begin{figure}[htbp]
\begin{flushleft}
\includegraphics[width=10.0cm,angle=-0]{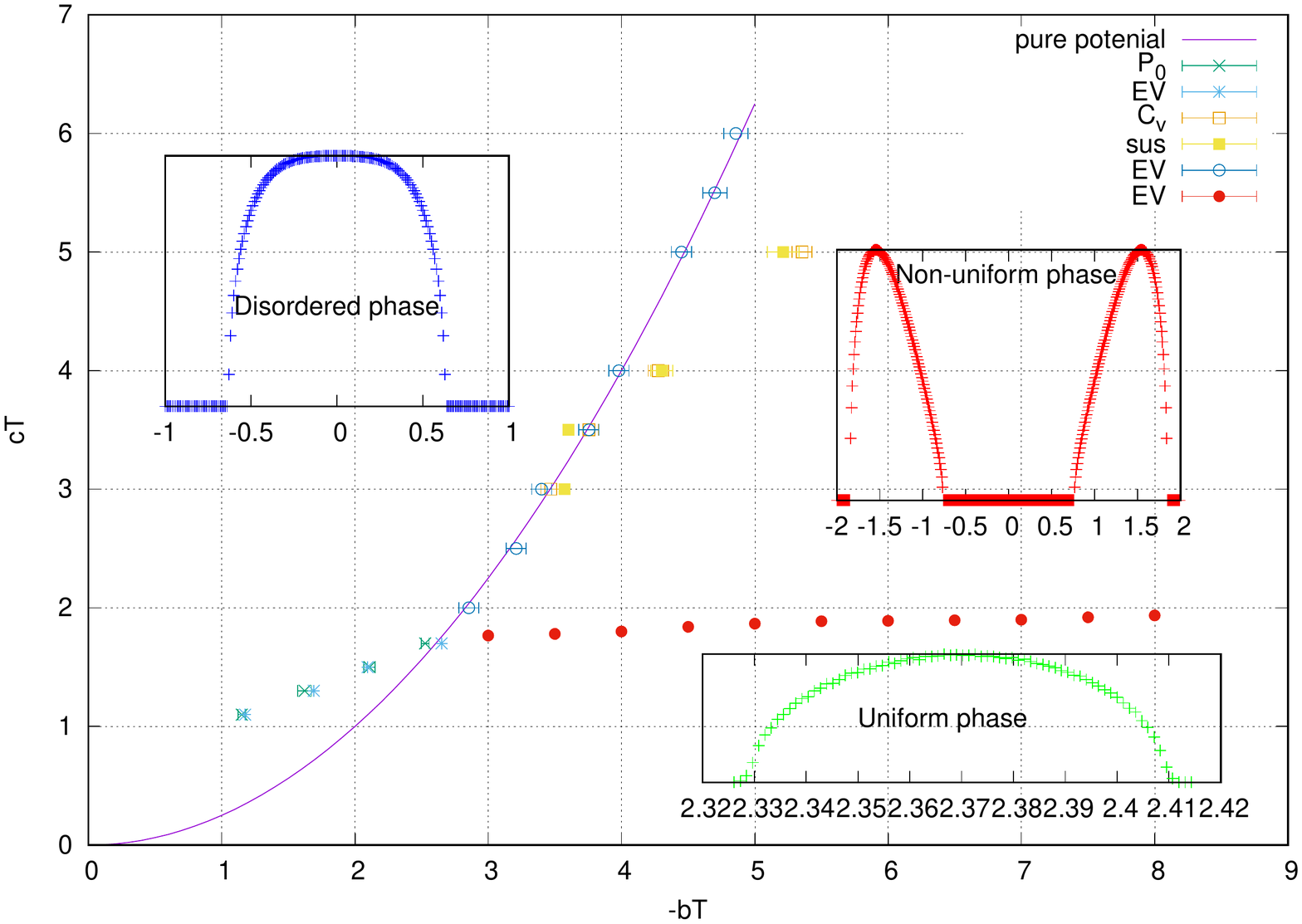}
\caption{The phase diagram of the noncommutative-phi four theory (\ref{matrix})+(\ref{phi-four}) on the fuzzy sphere as approximated by the multitrace matrix model (\ref{multitrace}). }\label{phase_diagram}
\end{flushleft}
\end{figure}

We have for both cases, i.e. for the  noncommutative phi-four theory (\ref{matrix})+(\ref{phi-four}) and for the random multitrace matrix model (\ref{multitrace}), the following phase structure:
\begin{enumerate}
\item A $2$nd order phase transition between disordered ($\Phi\sim 0$) and uniform-ordered ($\Phi\sim 1$) phases at small values of the quartic coupling. This is the usual Ising phase transition \cite{Onsager:1943jn}. The eigenvalue distribution $\rho(\lambda)$ as it transits between the  disordered and uniform-ordered phases in the multitrace matrix model (\ref{multitrace}) is shown in figure (\ref{commutative}).
\item A $2$nd order phase transition between non-uniform-ordered ($\Phi\sim \gamma$ with $\gamma^2=1$) and uniform-ordered ($\Phi\sim 1$) phases which is a continuation  of the Ising transition  to large values of the quartic coupling \cite{brazovkii}. The eigenvalue distribution $\rho(\lambda)$ as it transits between the  uniform-ordered and non-uniform-ordered phases in the multitrace matrix model (\ref{multitrace}) is shown in figure (\ref{noncommutative}).
\item A $3$rd order phase transition between disordered and non-uniform ordered phases. This is the matrix phase transition observed previously in the real quartic matrix model (\ref{purematrix}) (model without kinetic term). The eigenvalue distribution $\rho(\lambda)$ as it transits between the  disordered and non-uniform-ordered phases in the multitrace matrix model (\ref{multitrace}) is shown in figure (\ref{matrix}).
\item The three co-existence curves intersect at a triple point. The full phase diagram of the multitrace matrix model (\ref{multitrace}) in the plane $(-\tilde{B},\tilde{C})$, as measured by the Monte Carlo method, is shown in figure (\ref{phase_diagram}) where the coexistence line $\tilde{C}=\tilde{B}_c^2/4$ of the real quartic matrix model (\ref{purematrix}) is also included for comparison. The agreement with the phase diagram of noncommutative phi-four (\ref{matrix})+(\ref{phi-four}) on the fuzzy sphere is extremely favorable.

The observables used in the Monte Carlo measurement of the phase diagram include, in addition to the eigenvalue (EV) distribution  $\rho(\lambda)$, the specific heat $C_v=\langle S^2\rangle-\langle S\rangle^2$ where $S\equiv V_{\rm trunc}$, the susceptibility (${\rm sus}$) defined by $\chi=\langle |{\rm Tr}M|^2\rangle-\langle |{\rm Tr}M|\rangle^2$  and the power in the zero mode $P_0=\langle({\rm Tr}M)^2/N^2\rangle$. We also measure the magnetization $m=\langle |{\rm Tr} M|\rangle$, the total power $P_T=\langle {\rm Tr}M^2\rangle/N$ and the quartic coupling $\langle {\rm Tr}M^4\rangle$. 
\end{enumerate}
Now, the existence of  the uniform-ordered phase and the Ising phase transition between this uniform-ordered phase and disordered phase signal as usual the spontaneous symmetry breaking of the discrete symmetry $\Phi\longrightarrow -\Phi$. The fundamental underlying hypothesis in this article is the statement that the existence of an Ising uniform-ordered phase in a "pure matrix model" (such as the multitrace matrix model (\ref {multitrace}) which does not come with a pre-defined Laplacian operator) is an unambiguous signal for a lurking underlying geometry. In other words, an emergent geometry transition in this scenario (obtained by allowing the value of the coefficient $C^{\prime}$ of the multitrace term to change) is identified with the existence of a stable uniform-ordered phase and a corresponding Ising transition for some values of  the coefficient $C^{\prime}$.

On the other hand, the existence of the non-uniform-ordered phase signals the spontaneous symmetry breaking of translation symmetry (see for example \cite{Mejia-Diaz:2014lza}) which is quite remarkable for two reasons. First,  this breaking is possible even in two dimensions in contrast to the Coleman-Mermin-Wagner theorem \cite{Mermin:1966fe,Coleman:1973ci} and it is due to the fact that noncommutative field theories are non-local by construction. Second, this breaking can be extended to supersymmetric models allowing us to obtain spontaneous supersymmetry breaking which is usually quite hard to obtain otherwise (see \cite{Volkholz:2007kva} for a courageous attempt). The non-uniform-ordered phase is precisely the so-called stripe phase and it is the non-perturbative manifestation of the celebrated phenomena of the UV-IR mixing in noncommutative field theories \cite{Minwalla:1999px}.

\begin{figure}[htbp]
\begin{center}
\includegraphics[width=10.0cm,angle=-0]{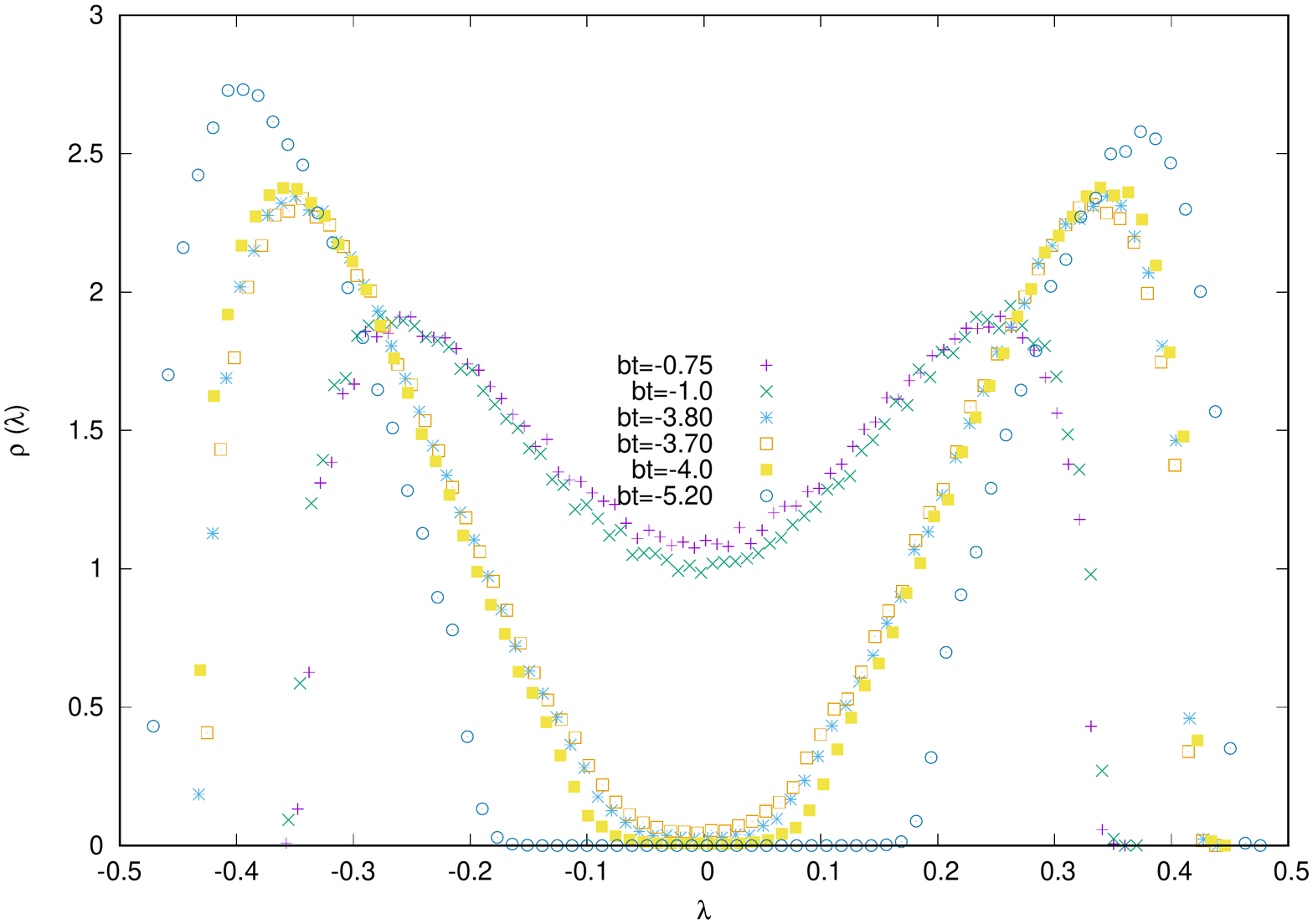}
\caption{The eigenvalue distribution as it transits between the disordered and non-uniform-ordered phases  in the multitrace matrix model (\ref{multitrace}). }\label{matrix}
\end{center}
\end{figure}

\section{Wilsonian renormalization group equation}
The partition function of the multitrace matrix model can be rewritten in the form

               \begin{eqnarray}
 Z&=&\int {\cal D}M~\exp(-N Tr_{N} V(M))\nonumber\\
 V(M)&=&\frac{g_2}{2}M^2+\frac{g_4}{4}M^4+\frac{g}{3N}(Tr_NM)M^3.
               \end{eqnarray}
               We have 
               \begin{eqnarray}
                 g_2=\pm 1~,~g_4=\frac{\tilde{C}}{\tilde{B}^2}~,~g=\frac{3}{4}\frac{\tilde{D}}{{\tilde{B}^2}}.
               \end{eqnarray}
               We choose $g_2=-1$ since $B$ is taken to be negative which is the region of interest for noncommutative scalar phi-four theories. Note also that the multitrace term is of the same order as the quartic term and stability requires that $\tilde{C}>-\tilde{D}$ which was noted previously on several occasions.

Now, by employing repeatedly large $N$ factorization of multi-point functions  of $U(N)-$invariant objects into product of one-point functions \cite{Higuchi:1994rv,Higuchi:1993nq,Higuchi:1993tg,Higuchi:1994dv} we can effectively convert multitrace terms into singletrace terms. In addition, by expanding the multitrace and quartic terms and using the Schwinger-Dyson identities in the cubic potential we can rewrite the above multitrace matrix model as a cubic potential of the form \cite{Ydri:2020efr}
                \begin{eqnarray}                 
                 &&V(M)=\frac{g_2^{\prime}}{2}M^2+\frac{g_3}{3}M^3\nonumber\\
                 &&g_2^{\prime}=\pm\sqrt{(g_2+\frac{g_4}{g_3^2})^2-2g_4}\nonumber\\
                 &&g_3=ga_1=\frac{g}{N}\langle {\rm Tr}M\rangle.
  \end{eqnarray}
  This is essentially a mean-field-approximation which seems to be exact in the large $N$ limit. Remark that the cubic coupling is proportional to the magnetization, i.e. $a_1\equiv m$.
  
\begin{figure}[htbp]
\begin{center}  
  \includegraphics[width=9cm,angle=-0]{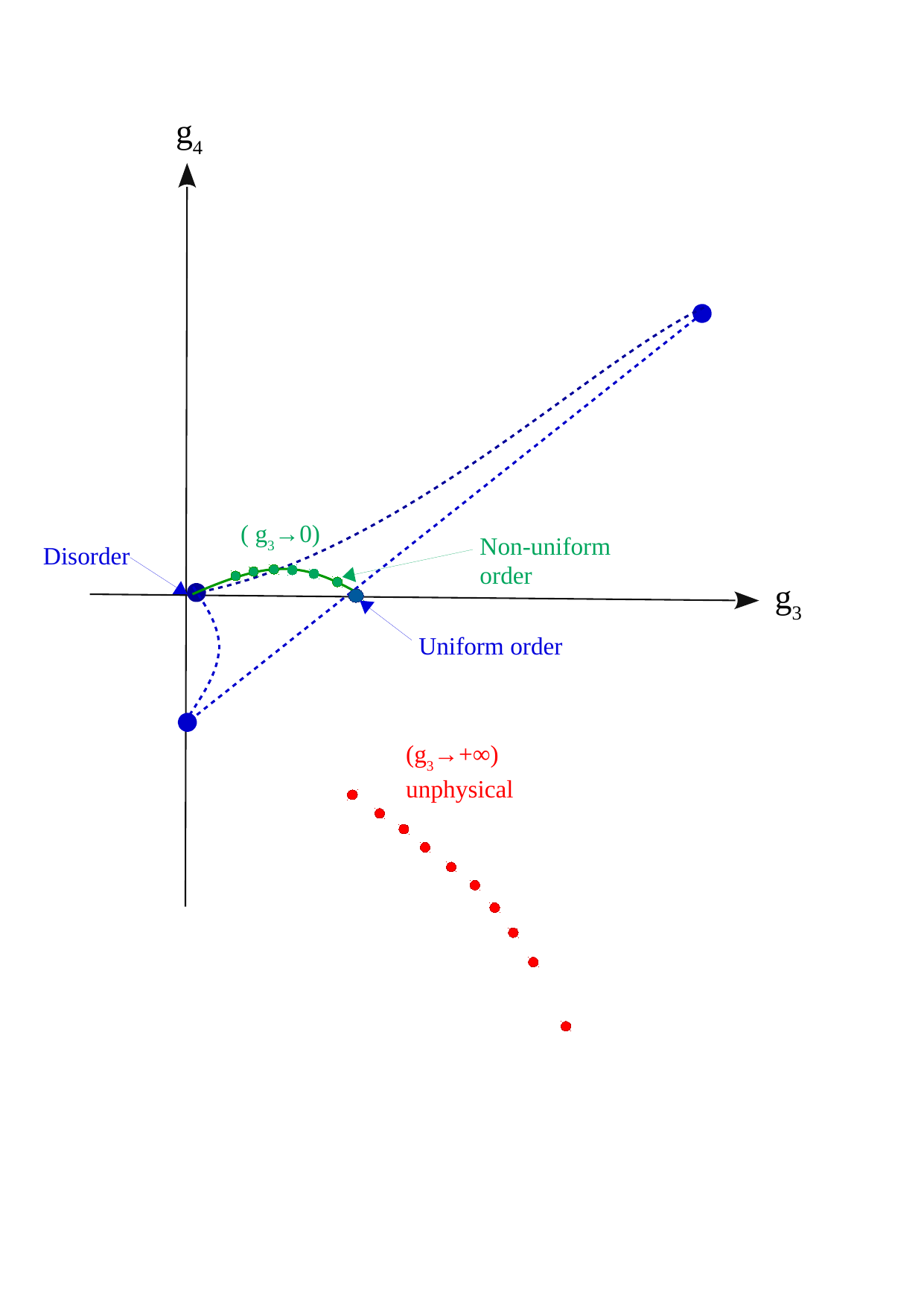}
\end{center}
\caption{The RG fixed points of the multitrace matrix model (\ref{multitrace}) (green and red lines) compared with the RG fixed points of the singletrace cubic-quartic matrix model (blue line)  \cite{Ydri:2020efr}.}\label{sketch2}
\end{figure}

  This cubic potential admits the standard Liouville quantum gravity fixed point  \cite{Higuchi:1994rv,Higuchi:1993nq,Higuchi:1993tg,Higuchi:1994dv}
  \begin{eqnarray}                 
                 \frac{g_3}{(g_{2*}^{\prime})^{\frac{3}{2}}}=\epsilon\equiv\frac{1}{432^{1/4}}.
  \end{eqnarray}
  The solution which goes through the two standard fixed points $(0,0)$ and  $(0,\epsilon)$ of the pure cubic potential is given explicitly by  \cite{Ydri:2020efr}
  \begin{eqnarray}                 
                 g_{4*}=g_3^2\bigg[g_3^2-\sqrt{g_3^4-2g_2g_3^2+\big(\frac{g_3}{\epsilon}\big)^{4/3}}-g_2\bigg].\label{result}
  \end{eqnarray}
  The behavior for $g_3^2\longrightarrow \infty$ and $g_3^2\longrightarrow 0$ is given explicitly by
   \begin{eqnarray}                 
                 g_{4*}=-\frac{1}{2}(\frac{g_3}{\epsilon})^{4/3}~,~g_3^2\longrightarrow \infty.
   \end{eqnarray}
   And
   \begin{eqnarray}                 
                 g_{4*}=g_3^2(-g_2-(\frac{g_3}{\epsilon})^{2/3})~,~g_3^2\longrightarrow 0.
   \end{eqnarray}
   We have then a critical line of fixed points (\ref{result}) which interpolate smoothly between the fixed point $(0,0)$ (interpreted as the fixed point of the $3$rd order matrix  coexistence line)  and  the fixed point $(0,\epsilon)$  (interpreted as the fixed point of the $2$nd order commutative/Ising  coexistence line)  going through a maximum in between. This critical line, restricted to the positive quadrant since the behavior $g_{4*}\longrightarrow -\infty$ as $g_3\longrightarrow + \infty$ is unphysical for noncommutative field theory, is interpreted as the critical line of fixed points associated with the noncommutative/stripe coexistence line. The RG fixed points are sketched on figure (\ref{sketch2}).

\section{Emergent geometry as emergence of the uniform-order}

To summarize, the non-uniform-ordered phase is important for noncommutative geometry and matrix models whereas the uniform-ordered phase  is essential to commutative geometry. In fact, the central proposal of this article is to turn the original logic of noncommutative geometry and its matrix models on its head by starting from matrix models and attempt to reach noncommutative geometry and not the other way around. This will be precisely/explicitly done by searching in the phase diagram for a uniform-ordered phase.

Hence, we will take as first principle the random multitrace matrix model (\ref{multitrace}) then attempt to reach the noncommutative field theory (\ref{matrix})+(\ref{phi-four}) by searching for a uniform-ordered phase in the phase diagram of the former.  This relies on three facts. First, multitrace matrix models such as  (\ref{multitrace}) do not involve in their definition a spectral triple specifying the geometry like those implicitly defining noncommutative field theories such as (\ref{matrix})+(\ref{phi-four}). Second, the uniform-ordered phase is a commutative order requiring the existence of an underlying space which is a priori commutative. Third, the underlying geometry turns out to be quantized, i.e. noncommutative geometry because of the existence of a triple point in which a commutative order (uniform/Ising), a noncommutative order (non-uniform/stripe) and a matrix order coexist. 

Thus, from the existence of a uniform-ordered phase  and a corresponding Ising phase transition, for some values of  the coefficient  of the multitrace coupling, we can infer the existence of an underlying space and an emergent geometry transition. The dimension of this emergent space can be determined from the critical exponents of the uniform-to-disordered phase transition by virtue of scaling and universality properties of second order phase transitions \cite{Wilson:1973jj}. This exercise was done in great detail in \cite{Ydri:2015zsa}.

Furthermore, the existence of a uniform-ordered phase and an Ising transition from a disordered phase to this uniform-ordered phase in a pure matrix model such as the multitrace matrix model (\ref{multitrace}) is indicative that this multitrace matrix model falls (for negative values of the multitrace coupling $C^{\prime}$) in the universality class of the noncommutative phi-four theory (\ref{matrix})+(\ref{phi-four}).  In other words, the multitrace matrix model (\ref{multitrace}) captures the same geometry as the geometry specified by the spectral triple which went implicitly into the definition of the noncommutative phi-four theory (\ref{matrix})+(\ref{phi-four}). For positive values of the  multitrace coupling $C^{\prime}$ the multitrace matrix model (\ref{multitrace}) does not sustain a uniform-ordered phase and hence it falls in the universality class of the real quartic matrix model (\ref{purematrix}). We have then an emergent geometry as we vary $C^{\prime}$ from positive to negative values (see figure (\ref{emergent_geometry})). The dimension of the underlying space, or more precisely this emergent geometry, are determined as we have seen from the measurement/calculation of the critical exponents characterizing the Ising transition.

Hence, it is the statistical physics of the commutative (uniform/Ising) phase of the multitrace matrix model that captures the geometry.  And, as it turns out,   this geometry is quantum in the sense that it is emergent (reminiscent of second quantization of geometry) but it is also quantized in the sense that it is noncommutative (first quantization of geometry).

Indeed, this emergent quantum geometry is necessarily noncommutative since the multitrace matrix model (\ref{multitrace}) includes necessarily a stripe or non-uniform-ordered phase which is a noncommutative phase by excellence (a transition from a disordered phase to a non-uniform-ordered phase is necessarily generated by the  single trace terms of the multitrace matrix model (\ref{multitrace})). This important fact can be further confirmed by studying the transition from non-uniform-ordered to uniform-ordered and verifying that this transition is also $2$nd order (similarly to the Ising transition) and it is in fact a continuation of the Ising transition to larger values of the quartic coupling.  

The precise metric on the emergent geometry can be fixed by studying the Wigner's semi-circle law near the origin, i.e. by studying the eigenvalue distribution of the matrix $M$ for vanishingly small values of the quartic coupling. This eigenvalue distribution captures clearly the properties of the free propagator.

This fourth ingredient in our proposal allows us, for example if the dimension is determined to be two from the critical exponents, to discriminate  between the noncommutative torus $\mathbb{T}^2_{\theta}$ and the fuzzy sphere $\mathbb{S}^2_N$ which can be both used to regularize non-perturbatively the Moyal-Weyl plane. In fact, these two spaces lead to different behavior of the radius $ \delta$ of the Wigner's semi-circle law as a function of the matrix size $N$  and the mass parameter $B$.

The Wigner's semi-circle law is obtained of course for the real quartic matrix model (\ref{purematrix}) with $C=0$, i.e. for the free model $B{\rm Tr} M^2$ with a squared radius given simply by
.\begin{eqnarray}
\delta^2=\frac{2N}{B}.\label{pred0}
\end{eqnarray}
As it turns out, a free noncommutative scalar field theory with mass parameter $m^2$ (the theory given by (\ref{matrix})+(\ref{phi-four}) with zero interaction) is itself characterized by a Wigner's semi-circle law  in stark contrast to free commutative scalar field theory. This stems from the fact that planar diagrams dominate over the non-planar ones in the limit of infinite cutoff \cite{Steinacker:2005wj,Nair:2011ux}. The Wigner semi-circle law in free noncommutative scalar field theory is given explicitly by \cite{Steinacker:2005wj}
\begin{eqnarray}
\rho(x)=\frac{2}{\pi\delta^2}\sqrt{\delta^2-x^2}.\label{pred1}
\end{eqnarray} 
By using a sharp UV  cutoff $\Lambda$ we compute in two dimensions the squared-radius

\begin{eqnarray}
\delta^2(m,\Lambda)=\frac{1}{\pi}\ln(1+\frac{\Lambda^2}{m^2}).\label{pred2}
\end{eqnarray}
This is indeed the behavior found on the fuzzy sphere $\mathbb{S}^2_N$ with $\Lambda=N/R$ and $m^2=b/aR^2$, i.e. $\Lambda^2/m^2=2\pi N/B$. The behavior of the squared-radius $\delta^2$ on the noncommutative torus $\mathbb{T}^2_{\theta}$ (with cutoff $\Lambda=\sqrt{{N\pi}/{\theta}}$) is found to be different from the above sharp UV cutoff result (\ref{pred2}) due to the different behavior of the propagator for large momenta.

\begin{figure}[htbp]
\begin{center}
\includegraphics[width=10cm,angle=-0]{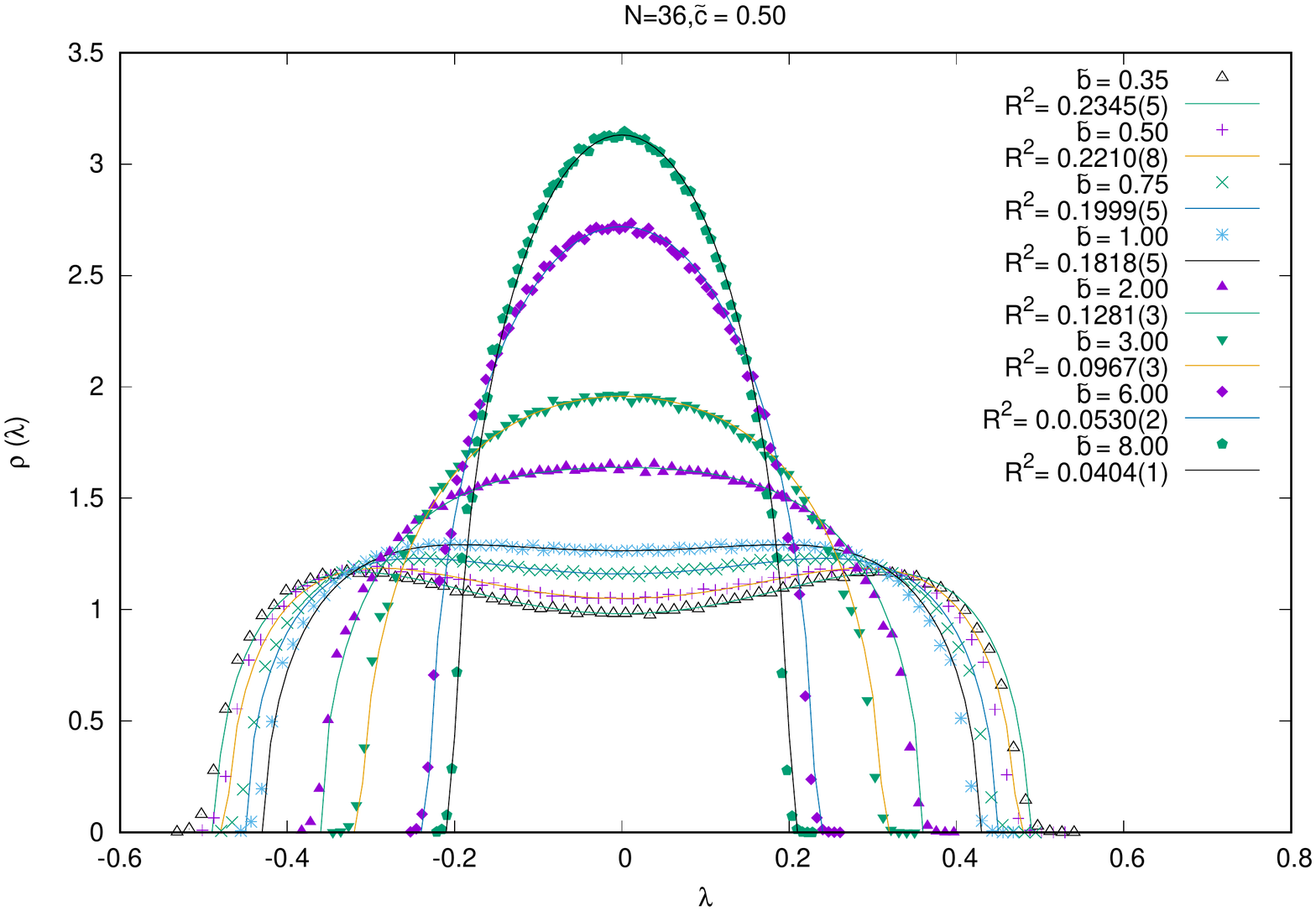} 
  \includegraphics[width=10cm,angle=-0]{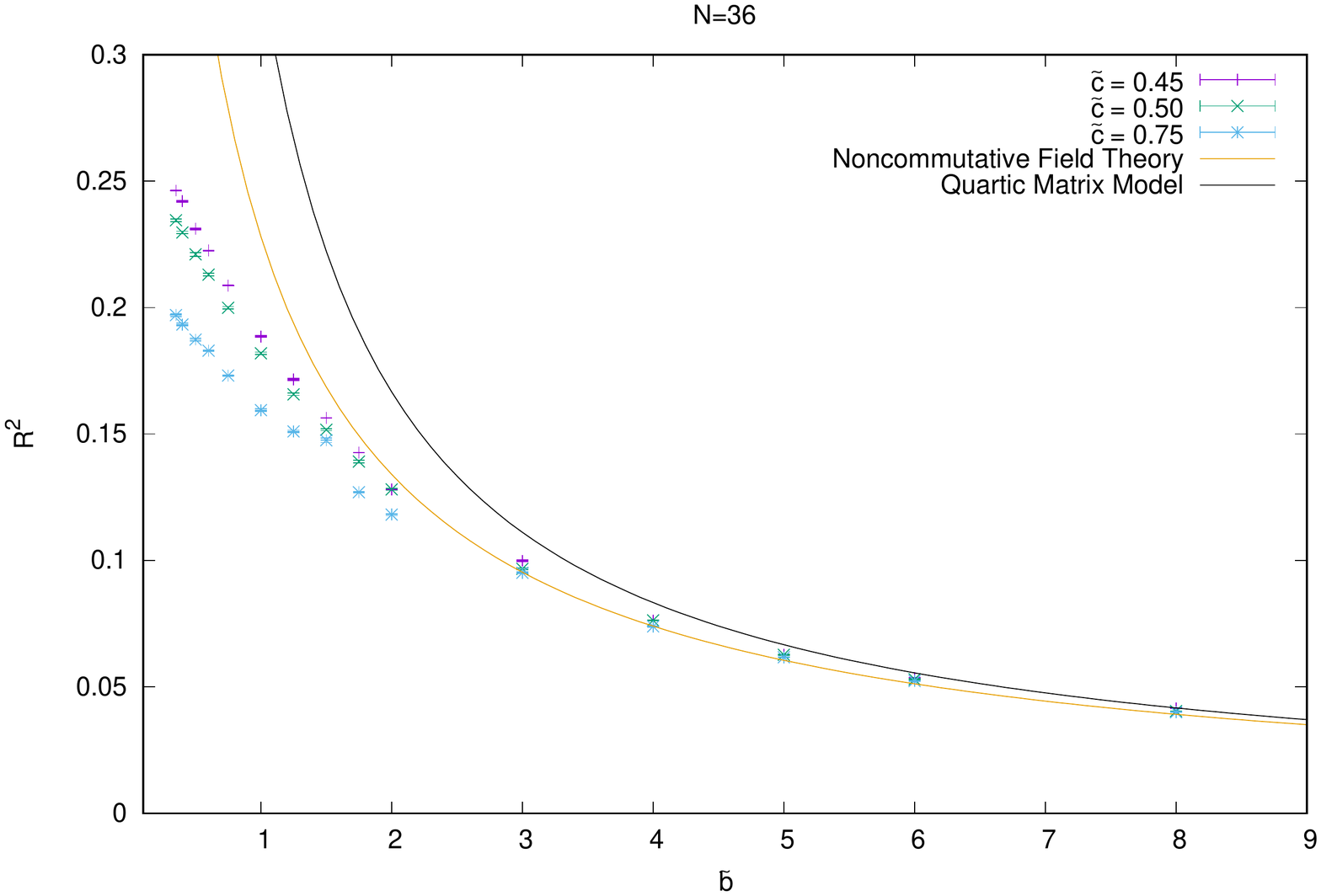}
\end{center}
\caption{Wigner's semicircle law.}\label{wigner}
\end{figure}

The Wigner's semi-circle law computed near the origin using the multitrace matrix model (\ref{multitrace}) should then be compared with the noncommutative field theory prediction (\ref{pred1})+(\ref{pred2}) for negative values of the multitrace coupling $C^{\prime}$. But for positive values of $C^{\prime}$ the behavior should be compared with the prediction of pure matrix model (\ref{pred0}). This will allow us to infer whether or not the underlying space or emergent geometry is indeed a fuzzy sphere $\mathbb{S}^2_N$ or a noncommutative torus $\mathbb{T}^2_{\theta}$. See figure (\ref{wigner}). But for more detail see \cite{Ydri:2015zsa}.

The fifth and final ingredient in our emergent geometry proposal consists in verifying explicitly the geometrical content of the multitrace matrix model (\ref{multitrace}) by expanding the matrix $M$ around the uniform-ordered configuration $M_0=m {\bf 1}$. We obtain an $SO(3)-$symmetric three-matrix model with a Chern-Simons term proportional to the value $m$ of the order parameter in the uniform-ordered phase, i.e. to the magnetization. This three-matrix model describes a noncommutative gauge theory on the fuzzy sphere \cite{Steinacker:2003sd}. The Chern-Simons term is precisely Myers term responsible for the condensation of the geometry \cite{Myers:1999ps,Azuma:2004zq}.

In summary, our proposal for "emergent geometry from random multitrace matrix models" consists of five ingredients  \cite{Ydri:2020efr,Ydri:2017riq,Ydri:2016daf,Ydri:2015zsa}:
\begin{enumerate}
\item The random multitrace matrix model (\ref{multitrace}) is taken as  "first principle". There is no Laplacian entering the definition of this action and thus there is no geometry a priori. First, we need to determine whether or not this pure matrix model contains in its phase diagram a uniform-ordered phase and an Ising transition from this uniform-ordered phase to disordered phase.
\item The existence of a uniform-ordered phase in a multitrace matrix model such as (\ref{multitrace}) signals the existence of an underlying space and an emergent geometry as we vary the multitrace coupling $C^{\prime}$ from positive to negative values. By using scaling and universality of second order phase transitions we can determine the dimension of the underlying space or emergent geometry from the critical exponents of the Ising line.
\item From the existence of a stripe or non-uniform-ordered phase with a transition line between uniform-ordered and non-uniform-ordered phases which is second order and a continuation of the Ising line we can infer that the underlying space or emergent geometry is in fact noncommutative.
\item The study of the Wigner's semi-circle law  of the multitrace matrix model (\ref{multitrace}) at the origin will allow us to discriminate decisively between the behavior of the noncommutative field theory and the real quartic matrix model.
\item In some interesting cases the expansion around the uniform-ordered phase exhibits the geometrical content of the multitrace matrix model as a gauge theory on a noncommutative background.
\end{enumerate}

\section{Conclusion}

Quantized geometry without Connes spectral triple of noncommutative geometry (first quantization of geometry) and without Yang-Mills matrix models of string theory (second quantization of geometry or quantization of noncommutative geometry relevant to quantum gravity) can be obtained by means of random multitrace matrix models. This proposal should be thought of as a generalization of the idea of discretization of random Riemannian surfaces with regular polygons (two-dimensional quantum gravity or $d=0$  bosonic string theory).

In this scenario the statistical physics of the random matrix model, in particular the properties of the commutative uniform-ordered phase more than the properties of the noncommutative non-uniform-ordered phase, play a crucial role in determining geometrical properties.

Furthermore, in this scenario of "emergent geometry through random multitrace matrix models" both the structure and the symmetry of the various multitrace couplings of the one-matrix model  are what underlies the dynamical process of quantum geometry, i.e. emergent geometry and topology change. This should be contrasted with the scenario of "emergent geometry through Yang-Mills matrix models" in which the structure and symmetry of the various singletrace terms of the many-matrix model is what underlies the dynamical process of quantum geometry \cite{Delgadillo-Blando:2007mqd,Ydri:2016kua}. 

In summary, a new scenario for quantized geometry is proposed in this article. In this scenario quantum geometry can emerge from "one-matrix multitrace scalar matrix models"  by probing the statistical physics of commutative phases of matter which should be contrasted with the usual scenario in which quantum geometry emerges from "many-matrix singletrace Yang-Mills matrix models" by probing the statistical physics of noncommutative phases of gauge theory. In both cases the geometry emerges as the system is cooled down in a well defined sense.

Quantum geometry is a fundamental ingredient of quantum gravity and in this novel scenario it appears in the form of a transition between the two phase diagrams of the real quartic matrix model and the noncommutative scalar phi-four field theory in which it is the commutative (and not the noncommutative) phase which plays the essential role. The matrix transition line in the real quartic matrix model is replaced as the geometry emerges by a matrix, noncommutative (stripe) and commutative (Ising) transition lines intersecting at a triple point.

The multitrace matrix model considered here is obtained by adding the term $C^{\prime}{\rm Tr} M {\rm Tr}M^3$ to the real quartic matrix model, i.e. there is no Laplacian operator (underlying metric) in this model. Yet, a noncommutative geometry emerges as we vary the multitrace coupling from positive to negative values. The multitrace coupling $C^{'}$ acts therefore as the temperature in this case playing in fact an analogous role to the role played by the gauge coupling constant in Yang-Mills matrix models.

Our proposal for emergent geometry from multitrace matrix models consists of five ingredients. First, we search for the geometry by searching in the phase diagram for an Ising uniform phase.  The dimension of the emergent geometry is determined from the value of the critical exponents of the Ising line whereas the metric is determined from the behavior of the Wigner's semi-circle law near the origin which allows us to discriminate between the behavior of the noncommutative field theory and the real quartic matrix model. The further existence of a non-uniform stripe order is indicative that the emergent geometry is noncommuting. Finally, in some important cases the expansion around the uniform-ordered phase exhibits explicitly the geometrical content of a gauge theory on a noncommutative background.

We have given arguments in support of this scenario from the large $N$ saddle point equation, from Monte Carlo simulation and from the renormalization group equation. 

Generalization to the multitrace matrix models of fuzzy projective spaces $\mathbb{CP}^n_N$ is straightforward \cite{Balachandran:2001dd,Saemann:2010bw}. In particular, the four-dimensional projective space $\mathbb{CP}^2_N$ is of a clear interest \cite{Alexanian:2001qj}. A much more challenging task is to generalize the multitrace approach to Lorentzian Poisson manifolds such as the noncommutative ${\rm AdS}^2_{\theta}$ \cite{Ho:2000fy,Ho:2000br,Jurman:2013ota,Pinzul:2017wch}. Some preliminary hints can be found in  \cite{Ydri:2021vxs}.


\begin{thebibliography}{99}

\bibitem{connes}
A. Connes, {\it Noncommutative Geometry}, Academic Press, London,1994. 

\bibitem{Connes:1996gi} 
  A.~Connes,
  ``Gravity coupled with matter and foundation of noncommutative geometry,''
  Commun.\ Math.\ Phys.\  {\bf 182}, 155 (1996).

\bibitem{Seiberg:1999vs}
N.~Seiberg and E.~Witten,
``String theory and noncommutative geometry,''
JHEP \textbf{09}, 032 (1999)
[arXiv:hep-th/9908142 [hep-th]].


\bibitem{Connes:1997cr}
A.~Connes, M.~R.~Douglas and A.~S.~Schwarz,
``Noncommutative geometry and matrix theory: Compactification on tori,''
JHEP \textbf{02}, 003 (1998)
[arXiv:hep-th/9711162 [hep-th]].



\bibitem{weyl}
  H.~Weyl,
  ``The Theory of Groups and Quantum Mechanics,''
  Dover, New York (1931).


\bibitem{Moyal:1949skv2}
  J.~E.~Moyal,
  ``Quantum mechanics as a statistical theory,''
  Proc.\ Cambridge Phil.\ Soc.\  {\bf 45}, 99 (1949).


\bibitem{Groenewold:1946kpv2}
  H.~J.~Groenewold,
  ``On the Principles of elementary quantum mechanics,''
  Physica {\bf 12}, 405 (1946).


\bibitem{Filk:1996dm}
T.~Filk,
``Divergencies in a field theory on quantum space,''
Phys. Lett. B \textbf{376}, 53-58 (1996)


\bibitem{Minwalla:1999px}
S.~Minwalla, M.~Van Raamsdonk and N.~Seiberg,
``Noncommutative perturbative dynamics,''
JHEP \textbf{02}, 020 (2000)
[arXiv:hep-th/9912072 [hep-th]].


\bibitem{Lee:2010zf1} 
  J.~Lee and H.~S.~Yang,
  ``Quantum Gravity from Noncommutative Spacetime,''
  J.\ Korean Phys.\ Soc.\  {\bf 65}, 1754 (2014)
  [arXiv:1004.0745 [hep-th]].


\bibitem{Blaschke:2010ye}
D.~N.~Blaschke and H.~Steinacker,
``Schwarzschild Geometry Emerging from Matrix Models,''
Class. Quant. Grav. \textbf{27}, 185020 (2010)
[arXiv:1005.0499 [hep-th]].


\bibitem{FroehlichGawedzki}
J.~Fr\"{o}hlich and  K.~Gaw\c{e}dzki,
{\sl Conformal Field Theory and Geometry of Strings},
Lectures given at Mathematical Quantum Theory Conference, 
Vancouver, Canada, 4-8 Aug 1993. 
Published in Vancouver 1993, 
Proceedings, Mathematical quantum theory, {\bf Vol. 1} 57-97,
[arXiv:hep-th/9310187].


\bibitem{rieffel}
M.~A.~Rieffel,
C*-algebras associated with irrational rotations,
Pacific J.Math. 93 (1981) 415-429.


\bibitem{Ambjorn:2000cs}
J.~Ambjorn, Y.~M.~Makeenko, J.~Nishimura and R.~J.~Szabo,
``Lattice gauge fields and discrete noncommutative Yang-Mills theory,''
JHEP \textbf{05}, 023 (2000)
[arXiv:hep-th/0004147 [hep-th]].

\bibitem{Hoppe} 
J. Hoppe, MIT Ph.D. Thesis, (1982). 


\bibitem{Madore:1991bw} 
  J.~Madore,
  ``The Fuzzy sphere,''
  Class.\ Quant.\ Grav.\  {\bf 9}, 69 (1992).

\bibitem{GarciaFlores:2009hf}
F.~Garcia Flores, X.~Martin and D.~O'Connor,
``Simulation of a scalar field on a fuzzy sphere,''
Int. J. Mod. Phys. A \textbf{24}, 3917-3944 (2009)
[arXiv:0903.1986 [hep-lat]].


\bibitem{Martin:2004un}
X.~Martin,
``A Matrix phase for the phi**4 scalar field on the fuzzy sphere,''
JHEP \textbf{04}, 077 (2004)
[arXiv:hep-th/0402230 [hep-th]].


\bibitem{Panero:2006bx}
M.~Panero,
``Numerical simulations of a non-commutative theory: The Scalar model on the fuzzy sphere,''
JHEP \textbf{05}, 082 (2007)
[arXiv:hep-th/0608202 [hep-th]].

\bibitem{Das:2007gm}
C.~R.~Das, S.~Digal and T.~R.~Govindarajan,
``Finite temperature phase transition of a single scalar field on a fuzzy sphere,''
Mod. Phys. Lett. A \textbf{23}, 1781-1791 (2008)
[arXiv:0706.0695 [hep-th]].



\bibitem{O'Connor:2007ea} 
  D.~O'Connor and C.~Saemann,
  ``Fuzzy Scalar Field Theory as a Multitrace Matrix Model,''
  JHEP {\bf 0708}, 066 (2007)
  [arXiv:0706.2493 [hep-th]].

\bibitem{Saemann:2010bw} 
  C.~Saemann,
  ``The Multitrace Matrix Model of Scalar Field Theory on Fuzzy CP**n,''
  SIGMA {\bf 6}, 050 (2010)
  [arXiv:1003.4683 [hep-th]].



\bibitem{Polychronakos:2013nca} 
  A.~P.~Polychronakos,
  ``Effective action and phase transitions of scalar field on the fuzzy sphere,''
  arXiv:1306.6645 [hep-th].

\bibitem{Tekel:2014bta} 
  J.~Tekel,
  ``Uniform order phase and phase diagram of scalar field theory on fuzzy CP**n,''
  arXiv:1407.4061 [hep-th].

\bibitem{Nair:2011ux} 
  V.~P.~Nair, A.~P.~Polychronakos and J.~Tekel,
  ``Fuzzy spaces and new random matrix ensembles,''
  Phys.\ Rev.\ D {\bf 85}, 045021 (2012)
  [arXiv:1109.3349 [hep-th]].

\bibitem{Tekel:2013vz} 
  J.~Tekel,
  ``Random matrix approach to scalar fields on fuzzy spaces,''
  Phys.\ Rev.\ D {\bf 87}, no. 8, 085015 (2013)
  [arXiv:1301.2154 [hep-th]].


\bibitem{Tekel:2015uza} 
  J.~Tekel,
  ``Phase strucutre of fuzzy field theories and multitrace matrix models,''
  Acta Phys.\ Slov.\  {\bf 65}, 369 (2015)
  [arXiv:1512.00689 [hep-th]].

\bibitem{Tekel:2015zga} 
  J.~Tekel,
  ``Matrix model approximations of fuzzy scalar field theories and their phase diagrams,''
  JHEP {\bf 1512}, 176 (2015)
  [arXiv:1510.07496 [hep-th]].


\bibitem{Subjakova:2020haa} 
  M.~Subjakova and J.~Tekel,
  ``Multitrace matrix models of fuzzy field theories,''
  arXiv:2006.13577 [hep-th].



\bibitem{Steinacker:2005wj} 
  H.~Steinacker,
  ``A Non-perturbative approach to non-commutative scalar field theory,''
  JHEP {\bf 0503}, 075 (2005)
  [hep-th/0501174].


\bibitem{DiFrancesco:1993cyw} 
  P.~Di Francesco, P.~H.~Ginsparg and J.~Zinn-Justin,
  ``2-D Gravity and random matrices,''
  Phys.\ Rept.\  {\bf 254}, 1 (1995)
  [hep-th/9306153].

\bibitem{Zarembo:1998uk} 
  K.~L.~Zarembo and Y.~M.~Makeenko,
  ``An introduction to matrix superstring models,''
  Phys.\ Usp.\  {\bf 41}, 1 (1998)
  [Usp.\ Fiz.\ Nauk {\bf 168}, 3 (1998)].



\bibitem{Balachandran:2001dd}
A.~P.~Balachandran, B.~P.~Dolan, J.~H.~Lee, X.~Martin and D.~O'Connor,
``Fuzzy complex projective spaces and their star products,''
J. Geom. Phys. \textbf{43}, 184-204 (2002)
[arXiv:hep-th/0107099 [hep-th]].


\bibitem{Alexanian:2001qj}
G.~Alexanian, A.~P.~Balachandran, G.~Immirzi and B.~Ydri,
``Fuzzy CP**2,''
J. Geom. Phys. \textbf{42}, 28-53 (2002)
[arXiv:hep-th/0103023 [hep-th]].

\bibitem{Brezin:1977sv} 
  E.~Brezin, C.~Itzykson, G.~Parisi and J.~B.~Zuber,
  ``Planar Diagrams,''
  Commun.\ Math.\ Phys.\  {\bf 59}, 35 (1978).



\bibitem{Shimamune:1981qf} 
  Y.~Shimamune,
 ``On The Phase Structure Of Large N Matrix Models And Gauge Models,''
  Phys.\ Lett.\ B {\bf 108}, 407 (1982).




\bibitem{Bietenholz:2004as}
W.~Bietenholz, F.~Hofheinz and J.~Nishimura,
``On the relation between non-commutative field theories at theta = infinity and large N matrix field theories,''
JHEP \textbf{05}, 047 (2004)
doi:10.1088/1126-6708/2004/05/047
[arXiv:hep-th/0404179 [hep-th]].


\bibitem{Becchi:2003dg}
C.~Becchi, S.~Giusto and C.~Imbimbo,
``The Renormalization of noncommutative field theories in the limit of large noncommutativity,''
Nucl. Phys. B \textbf{664}, 371-399 (2003)
[arXiv:hep-th/0304159 [hep-th]].


\bibitem{Grosse:2003nw}
H.~Grosse and R.~Wulkenhaar,
``Renormalization of phi**4 theory on noncommutative R**2 in the matrix base,''
JHEP \textbf{12}, 019 (2003)
[arXiv:hep-th/0307017 [hep-th]].

\bibitem{Onsager:1943jn} 
  L.~Onsager,
  ``Crystal statistics. 1. A Two-dimensional model with an order disorder transition,''
  Phys.\ Rev.\  {\bf 65}, 117 (1944).




\bibitem{brazovkii}
  S.~A.~Brazovkii,
  ``Phase Transition of an Isotropic System to a Nonuniform State,''
  Zh. Eksp. Teor. Fiz {\bf 68}, (1975) 175-185.










\bibitem{Mejia-Diaz:2014lza}
H.~Mej\'\i{}a-D\'\i{}az, W.~Bietenholz and M.~Panero,
``The continuum phase diagram of the 2d non-commutative $\lambda \phi^4$ model,''
JHEP \textbf{10}, 056 (2014)
[arXiv:1403.3318 [hep-lat]].










\bibitem{Mermin:1966fe}
N.~D.~Mermin and H.~Wagner,
``Absence of ferromagnetism or antiferromagnetism in one-dimensional or two-dimensional isotropic Heisenberg models,''
Phys. Rev. Lett. \textbf{17}, 1133-1136 (1966)


\bibitem{Coleman:1973ci}
S.~R.~Coleman,
``There are no Goldstone bosons in two-dimensions,''
Commun. Math. Phys. \textbf{31}, 259-264 (1973)

\bibitem{Volkholz:2007kva}
J.~Volkholz and W.~Bietenholz,
``Simulations of a supersymmetry inspired model on a fuzzy sphere,''
PoS \textbf{LATTICE2007}, 283 (2007)
[arXiv:0808.2387 [hep-th]].



\bibitem{Wilson:1973jj}
K.~G.~Wilson and J.~B.~Kogut,
``The Renormalization group and the epsilon expansion,''
Phys. Rept. \textbf{12}, 75-199 (1974)


\bibitem{Steinacker:2003sd} 
  H.~Steinacker,
  ``Quantized gauge theory on the fuzzy sphere as random matrix model,''
  Nucl.\ Phys.\ B {\bf 679}, 66 (2004).
  [hep-th/0307075].





\bibitem{Myers:1999ps} 
  R.~C.~Myers,
  ``Dielectric branes,''
  JHEP {\bf 9912}, 022 (1999),
  [hep-th/9910053].

\bibitem{Azuma:2004zq}
  T.~Azuma, S.~Bal, K.~Nagao and J.~Nishimura,
  ``Nonperturbative studies of fuzzy spheres in a matrix model with the
  Chern-Simons term,''
  JHEP {\bf 0405} (2004) 005
  [arXiv:hep-th/0401038].


\bibitem{Higuchi:1994rv} 
  S.~Higuchi, C.~Itoi, S.~Nishigaki and N.~Sakai,
  ``Renormalization group flow in one and two matrix models,''
  Nucl.\ Phys.\ B {\bf 434}, 283 (1995)
  Erratum: [Nucl.\ Phys.\ B {\bf 441}, 405 (1995)]
  [hep-th/9409009]. See also \cite{Kawamoto:2013laa}.



  
\bibitem{Higuchi:1993nq} 
  S.~Higuchi, C.~Itoi, S.~Nishigaki and N.~Sakai,
  ``Nonlinear renormalization group equation for matrix models,''
  Phys.\ Lett.\ B {\bf 318}, 63 (1993)
  doi:10.1016/0370-2693(93)91785-L
  [hep-th/9307116].

  

\bibitem{Higuchi:1993tg} 
  S.~Higuchi, C.~Itoi, S.~Nishigaki and N.~Sakai,
  ``Renormalization group approach to discretized gravity,''
  hep-th/9307065.


\bibitem{Higuchi:1994dv} 
  S.~Higuchi, C.~Itoi, S.~Nishigaki and N.~Sakai,
  hep-th/9409157.


\bibitem{Ho:2000fy}
  P.~M.~Ho and M.~Li,
  ``Large N expansion from fuzzy AdS(2),''
  Nucl.\ Phys.\ B {\bf 590}, 198 (2000)
  [hep-th/0005268].



\bibitem{Ho:2000br}
  P.~M.~Ho and M.~Li,
  ``Fuzzy spheres in AdS / CFT correspondence and holography from noncommutativity,''
  Nucl.\ Phys.\ B {\bf 596}, 259 (2001)
  [hep-th/0004072].



\bibitem{Jurman:2013ota}
  D.~Jurman and H.~Steinacker,
  ``2D fuzzy Anti-de Sitter space from matrix models,''
  JHEP {\bf 1401}, 100 (2014)
  [arXiv:1309.1598 [hep-th]].



\bibitem{Pinzul:2017wch}
  A.~Pinzul and A.~Stern,
  ``Non-commutative $AdS_2/CFT_1$ duality: the case of massless scalar fields,''
  Phys.\ Rev.\ D {\bf 96}, no. 6, 066019 (2017)
  [arXiv:1707.04816 [hep-th]].




\bibitem{Ydri:2013zya}
B.~Ydri and R.~Ahmim,
``Matrix model fixed point of noncommutative $\phi^4$ theory,''
Phys. Rev. D \textbf{88}, no.10, 106001 (2013)
[arXiv:1304.7303 [hep-th]].


\bibitem{Ydri:2014rea}
B.~Ydri,
``New algorithm and phase diagram of noncommutative $\phi^4$ on the fuzzy sphere,''
JHEP \textbf{03}, 065 (2014)
[arXiv:1401.1529 [hep-th]].


\bibitem{Ydri:2015zsa}
B.~Ydri, A.~Rouag and K.~Ramda,
``Emergent geometry from random multitrace matrix models,''
Phys. Rev. D \textbf{93}, no.6, 065055 (2016)
[arXiv:1509.03572 [hep-th]].

\bibitem{Ydri:2017riq}
B.~Ydri, C.~Soudani and A.~Rouag,
``Quantum Gravity as a Multitrace Matrix Model,''
Int. J. Mod. Phys. A \textbf{32}, no.31, 1750180 (2017)
[arXiv:1706.07724 [hep-th]].


\bibitem{Ydri:2020efr}
B.~Ydri and R.~Ahmim,
``Wilsonian Matrix Renormalization Group,''
[arXiv:2008.09564 [hep-th]].





\bibitem{Ydri:2016daf}
B.~Ydri,
``The multitrace matrix model: An alternative to Connes NCG and IKKT model in 2 dimensions,''
Phys. Lett. B \textbf{763}, 161-163 (2016)
[arXiv:1608.02758 [hep-th]].







\bibitem{Delgadillo-Blando:2007mqd}
R.~Delgadillo-Blando, D.~O'Connor and B.~Ydri,
``Geometry in Transition: A Model of Emergent Geometry,''
Phys. Rev. Lett. \textbf{100}, 201601 (2008)
[arXiv:0712.3011 [hep-th]].

\bibitem{Ydri:2016kua}
B.~Ydri, R.~Khaled and R.~Ahlam,
``Geometry in transition in four dimensions: A model of emergent geometry in the early universe,''
Phys. Rev. D \textbf{94}, no.8, 085020 (2016)
[arXiv:1607.06761 [hep-th]].


\bibitem{Ydri:2021vxs}
B.~Ydri and L.~Bouraiou,
``The AdS\textasciicircum{}2\_\ensuremath{\theta}/CFT\_1 Correspondence and Noncommutative Geometry III: Phase Structure of the Noncommutative AdS\textasciicircum{}2\_\ensuremath{\theta} x S\textasciicircum{}2\_N,''
[arXiv:2109.01010 [hep-th]].

\end{thebibliography}
\end{document}